\documentclass[showpacs,11pt]{amsart}
\usepackage{amsfonts}
\usepackage{amsmath}
\usepackage{amssymb}
\usepackage{graphicx}%
\usepackage{graphicx}% Include figure files
\usepackage{bm}% bold math

\input{ueur.fd}
\DeclareMathVersion{euler}
\DeclareMathVersion{eulerbold}
\SetSymbolFont{letters}  {euler}{U}{eur} {m}{n}
\SetSymbolFont{letters}  {eulerbold}{U}{eur} {b}{n}
\SetSymbolFont{operators}{eulerbold}{OT1}{cmr} {bx}{n}
\SetSymbolFont{symbols}  {eulerbold}{OMS}{cmsy}{b}{n}
\SetMathAlphabet\mathsf{eulerbold}{OT1}{cmss}{bx}{n}
\SetMathAlphabet\mathit{eulerbold}{OT1}{cmr}{bx}{it}
\def\eulermath{\@nomath\eulermath
              \mathversion{euler}}
\def\uneulermath{\@nomath\uneulermath
              \mathversion{normal}}
\def\eulerbfmath{\@nomath\eulerbfmath
              \mathversion{eulerbold}}
\def\uneulerbfmath{\@nomath\uneulerbfmath
              \mathversion{normal}}

\let\newcommand=\providecommand
\makeatother

\begin{document}

\title
[Statistical analysis of the $\pi$--mode solution for the FPU $\beta$ system ]{Thermostatistics in the neighborhood of the $\pi$--mode solution for the Fermi--Pasta--Ulam $\beta$ system: from weak to strong chaos}
\author{Mario Leo, Rosario Antonio Leo}
\address{
Dipartimento di Fisica, Universit\`a del Salento, Via per Arnesano, 73100 -- Lecce, Italy}
\author{Piergiulio Tempesta}

\email{mario.leo@le.infn.it, leora@le.infn.it}
\address{Departamento de F\'{\i}sica Te\'{o}rica II, Facultad de F\'{\i}sicas, Ciudad Universitaria, Universidad
Complutense, 28040 -- Madrid, Spain}
\email{p.tempesta@fis.ucm.es}

\begin{abstract}
\noindent
We consider a $\pi$--mode solution of the Fermi--Pasta--Ulam $\beta$ system. By perturbing it, we study the system as a function of the energy density from a regime where the solution is stable to a regime, where is unstable, first weakly and then strongly chaotic. We introduce, as indicator of stochasticity, the ratio $\rho$ (when is defined) between the second and the first moment of a given probability distribution. We will show numerically that the transition between weak and strong chaos can be interpreted as the symmetry breaking of a set of suitable dynamical variables. Moreover, we show that in the region of weak chaos there is a numerical evidence that the thermostatistic is governed by the Tsallis distribution.

\end{abstract}
\keywords {Anharmonic lattices; Periodic solutions; Stability}
\vfill\eject
\date{March 24, 2010}
\maketitle

\tableofcontents

\section{Introduction}

Since its discovery, the celebrated Fermi--Pasta--Ulam (FPU) system \cite{fermi} has represented a paradigmatic model for the analysis of energy equipartition, stochastic resonances \cite{ruffo} and thermalization in nonlinear systems (for a recent account, see \cite{gallavotti}). In order to explain its rich phenomenology, several approaches have been proposed. Zabusky and Kruskal \cite{ZK}, by analyzing the string dynamics in the continuum limit, discovered solitary waves and started the modern theory of nonlinear integrable systems. Another approach, due to Izrailev and Chirikov \cite{isr} and many others, was addressed to the numerical determination of "stochasticity thresholds", that marked the transition from recurrences to thermalization and equilibrium.

In the last two decades it has been shown that many complex systems possess weakly chaotic regimes, such as those exhibiting long--range particle interactions, strong correlations, scale invariance, properties of multifractality, etc..  New physical phenomena are expected at the edge of chaos.

The approach first proposed in \cite{Tsallis1} aims to a generalization of the standard statistical mechanics. As is well known, the Boltzmann--Gibbs thermostatistics offers the natural theoretical framework to describe nonintegrable and fully chaotic dynamics. This yields eventually to ergodicity and mixing in phase space. A natural question is how to describe situations when the system exhibits a weakly chaotic behavior, the ergodic hypothesis typically is not verified and the statistical mechanics of Boltzmann and Gibbs (BG) fails to provide a correct theoretical framework. The classical picture is usually restored in the strongly chaotic regime. The approach, nowadays called \textit{nonextensive statistical mechanics} \cite{Tbook}, has been proposed in order to handle these more general situations, and in particular  deals with the case of power--law divergencies of the sensitivity to the initial conditions. At the heart of the theory there is a generalization of the Boltzmann--Gibbs entropy, the $S_q$ entropy, that depends on a real parameter, the entropic index $q$.

The literature on this topic has been increasing dramatically in the recent years (for a regularly updated bibliography, see \cite{bibl}).

Motivated by this current research, in this paper we analyze the statistical behaviour of the FPU $\beta$ system \cite{fermi} when a $\pi$--mode solution is initially excited. We describe both numerically and analytically, as a function of the energy density, the transition of the system from a \textit{stable} to a \textit{strongly chaotic regime} by following the time evolution of a suitable observable associated to the \textit{exact $\pi$--mode solution}. This observable physically corresponds to a geometric \textit{symmetry} of the system that is lost when the system is perturbed. The analysis of this observable offers a very accurate tool for the study of the evolution of the system.

An interesting result of our investigation is that in the weakly chaotic regime the model appears to be described by the Tsallis (q--Gaussian) statistics. Recent generalizations of the central limit theorem \cite{UTS}, \cite{UTG} provide a theoretical framework for the wide appearance of such statistics in Physics. These theorems claim that, under suitable hypotheses, $q$--Gaussian distributions should govern the weakly chaotic regimes, instead of Gaussian ones(see also \cite{TLSM}).

However, the fact that a Tsallis statistics seems to play an important r\^ole in the FPU $\beta$ model, at first sight is quite surprising, since we are dealing with a Hamiltonian system possessing a \textit{short--range interaction}, whereas, within the class of Hamiltonian systems, usually nonextensive regimes are observed in long--range interacting many--body systems.

The appearance of the Tsallis distribution for the FPU chain in the region of weak chaos could be a consequence of the choice of the initial condition to which an exact one-mode solution is associated. When a sufficiently small perturbation is applied, the occurrence of q-Gaussians can be expected (see, for instance, \cite{AT}, \cite{MPR}, \cite{TTB}). When the exact solution is further perturbed and the energy density is increased, the weakly--chaotic behaviour is replaced by a strongly chaotic one leading to ergodicity and to the classical Boltzmann--Gibbs statistics.

In order to detect accurately this transition, we introduce an \textit{indicator of stochasticity} $\rho$, that estimates the deviation of a generic assigned distribution from the Gaussian behaviour for any value of the excitation energy density. It is a function of the dynamical variables of the configuration space only.  The usefulness of the function $\rho$ relies on the fact that it is model--independent, since it can be used to characterize the behaviour of any complex system.

In Section 2, we review briefly some theoretical aspects of nonextensive thermostatistics. In Section 3, we propose our analysis of the FPU chain, from the initial conditions selected towards the strongly chaotic region of the phase space. In Section 4, the numerical results obtained are reported. In Section 5, some open problems related to our work are discussed.

\section{The nonextensive scenario}

Let us consider a system in classical statistical mechanics, whose associated probabilities are $p_{i}$ $(i=1,...,W)$, satisfying the condition $\sum_{i=1}^{W}p_i=1$. Here $W$ is the total number of possible (microscopic) configurations of the system. In \cite{Tsallis1}, the following entropy was introduced:
\begin{equation}
S_q=k\frac{1-\sum_{i=1}^{W} p_{i}^{q}}{q-1},\label{TE}
\end{equation}
where $q \in \mathbb{R}$ and $k$ is a positive constant. It is immediate to ascertain that it reduces, in the limit $q\rightarrow 1$, to the Boltzmann--Gibbs entropy:

\begin{equation}
S_1 = \lim_{q\rightarrow 1} S_q \equiv S_{BG} = -k \sum_{i=1}^{W}p_i \ln p_i. \label{B}
\end{equation}
If we introduce the $q$--exponential, defined by
\begin{equation}
e_q^x:=\left[1+(1-q)x\right]^{1/(1-q)},
\end{equation}
and the $q$--logarithm
\begin{equation}
\ln_q x:=\frac{x^{1-q}-1}{1-q},
\end{equation}
the entropy $S_q$ can be seen as a $q$--deformation of $S_{BG}$:
\begin{equation}
S_q = -k \sum_{i=1}^{W}p_i \ln_q p_i. \label{TS}
\end{equation}
The entropy $S_q$ possesses many physical properties. Two of them are particularly relevant.

a) \textit{Nonadditivity}. Given two probabilistically independent subsystems $A$ and $B$ of a given system, we have that
\begin{equation}
\frac{S_q (A + B)}{k}=\frac{S_q (A)}{k}+\frac{S_q (B)}{k}+(1-q)\frac{S_q (A)}{k}\frac{S_q (B)}{k} \label{nonadd}.
\end{equation}
Therefore, the entropy (\ref{TE}) is nonadditive, according to the definition proposed by Penrose \cite{Penrose}. In the literature, the cases $q<1$ and $q>1$ are usually referred to as super--additive and sub--additive, respectively.

b) \textit{Extensivity}. For systems with strictly or asymptotically scale--invariant correlations \cite{TGS}, \cite{RST} or global long--range interactions \cite{CT}, \textit{for special values of $q$}, $S_q$ satisfies the relation
\begin{equation}
S_q\sim N
\end{equation}
i.e. is proportional to the number of particles of the system. This is crucial, in order for a statistical mechanics to be a meaningful and widely applicable one, as already Clausius pointed out.

Unfortunately, the notions of additivity and extensivity have often been confused in the literature, and $S_q$, being nonadditive, has been referred to as nonxtensive as well. This use is indeed erroneous, although widespread in the literature. Von Neumann entropy, for instance, is additive but in general nonextensive. Both $S_{BG}$ and $S_q$ may or may not be extensive, depending on the specific physical system considered \cite{TGS}.

Tsallis entropy can be considered as the simplest nontrivial generalization of Boltzmann--Gibbs entropy: in addition to extensivity, it possesses all nice properties of the classical entropy (as concavity, Lesche--stability \cite{Lesche}, finiteness of entropy production for unit time, etc.), except additivity, which is replaced by the condition (\ref{nonadd}).
A crucial result has been obtained in several recent papers, as \cite{UTS}, \cite{UTG}, where $q$--extensions of the Central Limit Theorem have been proposed. In these works it has been shown that, when we deal with large sets of random variables with correlation (it may happen for instance in physical systems with a weakly chaotic regime), q--Gaussian probability distributions emerge as attractors, instead of Gaussian ones.
In the last twenty years, the nonextensive scenario has been widely investigated: many interesting physical systems at the edge of chaos, both in classical and quantum mechanics, have been shown to be conveniently described by the Tsallis statistics. Other relevant applications have been found in economics, linguistics, biosciences, social sciences, self--organized criticality, etc. \cite{bibl}. In the following, we will show how the FPU chain, under suitable conditions, admits in a specific region of the phase space a description in terms of the nonextensive statistics.

\section{The $\pi$--mode solution: a statistical analysis}

Let us now describe the main features of the FPU $\beta$ system with $N$ oscillators and periodic conditions. Let $x_i$ denote the displacement of the $i$--th particle of the nonlinear chain from its equilibrium position. The Hamiltonian of the model reads
\begin{equation}
H = \frac{1}{2} \sum_{i=1}^{N}{p_i}^2 + \frac{1}{2} \sum_{i=1}^{N} \left(x_{i+
1} - x_i \right)^2 + \frac{\beta}{4} \sum_{i=1}^{N} \left(x_{i+1} - x_i \right)^4
\label{eq:2}
\end{equation}
with
\begin{equation}
x_{N+1} = x_1 \qquad and \qquad \beta > 0.
\end{equation}
All quantities are dimensionless. If we introduce the normal coordinates $Q_i$ and $P_i$ of the normal mode through the relations
\begin{equation}
Q_i=\sum_{j=1}^{N}S_{ij}x_j \qquad P_i=\sum_{j=1}^{N}S_{ij}p_j,
\end{equation}
with
\begin{equation}
S_{ij}=\frac{1}{\sqrt{N}}\left(\sin \frac{2 \pi ij }{N} +\cos \frac{2 \pi ij }{N}\right),
\end{equation}
the harmonic energy of the mode $i$ is
\begin{equation}
E_i=\frac{1}{2}\left(P_i^2+\omega_i^2 Q_i^2\right),
\end{equation}
where for periodic boundary conditions
\begin{equation}
\omega_i^2=4 \sin^2 \frac{\pi i}{N} \label{puls}.
\end{equation}
For $\beta=0$, all normal modes oscillate independently and their energies $E_i$ are constant of the motion. In the anharmonic case $(\beta \neq 0)$, the normal modes are instead coupled, and the variables $Q$ have no longer simple sinusoidal oscillations.

Given a linear mode, if its excitation energy and the coupling nonlinear parameter are small, the energy exchange with the other modes also remains small and periodic. However, when the nonlinear effects become larger, a conspicuous exchange of energy among all normal modes is observed. In \cite{pet1}--\cite{CRZ}, the concept of strong stochasticity threshold (SST) has been introduced. It is defined as the energy density threshold that characterizes the transition of the system dynamics from weak to strong chaos during the relaxation of the system towards ergodicity and equipartition.

Besides, it is well known that, for a periodic FPU $\beta$ chain, there are nonlinear one--mode exact solutions (OMSs) ($\pi$--mode, $\pi/2$--mode, etc.) \cite{poggi} corresponding to the values of the integer mode number
\begin{equation}
n=\frac{N}{4}, ~~\frac{N}{3}, ~~\frac{N}{2}, ~~\frac{2}{3}N, ~~\frac{3}{4}N.
\end{equation}
If we excite one of these nonlinear modes and integrate the corresponding equations of the motion, the finite precision of the numerical algorithm naturally generates a perturbation of the mode. Beyond a certain threshold value $\epsilon_{t}$ of the energy density $\epsilon$, the nonlinear mode becomes unstable. In \cite{caf}--\cite{leo2}, this mechanism has been extensively used to analyze the stability properties of the $\pi$--mode and $\pi/2$--mode ($n = N/4$), both for positive and negative values of the nonlinearity parameter $\beta$.

What is the route towards equipartition, ergodicity or chaos when $\epsilon > \epsilon_{t}$?  Qualitatively, the behaviour of the system is the following. For $\epsilon > \epsilon_{t}$, the energy of the OMS is no longer constant. For small values of $\epsilon$ above the threshold, the $\pi$--mode loses and recovers almost completely its initial energy. In this recurrence region, if one increases $\epsilon$, only the fraction of the energy ceded to other modes and the period of recurrence change. For larger values of $\epsilon$, a more and more irregular behaviour is observed: the energy ceded increases and the periodicity of the recurrence is lost, while the system tends towards the equipartition of the whole initial energy.

A crucial point is the choice of indicators able to reveal the existence of thresholds. To this aim, several indicators have been introduced in the literature related, for instance, to the rate of energy exchange among normal modes, to geometrical properties of trajectories in phase space and to single--particle correlation functions. Collective spectral parameters have also been proposed \cite{livi1}. In particular, the normalized spectral entropy has been used to reveal the existence of a SST \cite{pet1}.

In this work, by using a new global indicator $\rho$,  we present, as a function of energy density, a statistical analysis of the FPU $\beta$ system, when the  $\pi$--mode is initially excited. This indicator is related to the distribution of the values of a physical observable which remains constant during the evolution of the system, if it is stable. As is well known, when one excites the $\pi$--mode, the variable $x_i$ is related to the modal variable $Q_{N/2}$ by
\begin{equation}
x_i(t)=\frac{1}{\sqrt{N}}(-1)^{i}Q_{N/2}(t) \label{qi}.
\end{equation}
Therefore, it is natural to introduce the observables
\begin{equation}
\eta_i=x_i+x_{i-1} \label{ei}.
\end{equation}
Indeed, the quantities $\eta_i$ are always equal to zero during the time evolution of the
system, if it is stable, independently of the choice of the initial condition $Q_{N/2}(0)$. Instead, when the energy density is greater than the instability threshold value $\epsilon_t$, the $\eta_i$ are no longer equal to zero. The distribution of the values of $\eta_i$ then depends on the exchange of energy among the $\pi$--mode and the other modes, rather than the statistic of the numerical integration errors. We will show numerically that the transition from weak to strong chaos can be interpreted as the breaking of the symmetry described by eqs. (\ref{qi}), (\ref{ei}). Taking into account these considerations, we introduce, as an indicator of stochasticity, the ratio between the second and the first moment of a given probability distribution
\begin{equation}
\rho = \frac{\sigma}{\theta},
\end{equation}
when they are defined and the first moment is not zero. In the case of symmetric distribution functions, $\theta$ is the mean value of the modulus of the differences between the values of the observable and its mean value.
In our analysis, we distinguish two possibilities.

a) The distribution is normal, i.e. described by the Gauss function
\begin{equation}
f(\xi) = \frac{a}{\sqrt{\pi}} ~\exp{(- a^{2} \xi^{2})},
\label{eq:5}
\end{equation}
\noindent where $a$ is a parameter. One has the theoretical value $\rho = \frac{\sigma}{\theta} = \sqrt {\frac{\pi}{2}}.$
This result is characteristic of normal distributions and is utilized to estimate if a distribution of measurements satisfies the Gauss distribution.

b) The distribution is a Tsallis distribution:
\begin{equation}
f(\xi) = a \left(1-(1-q) b^{2} \xi^{2}\right)^{\frac{1}{1-q}}.
\label{eq:81}
\end{equation}
\noindent with  $a$ and $q$ dependent on $\epsilon$,
\begin{equation}
b = a  \sqrt{\pi} ~\frac{\Gamma \left(\frac{3 - q}{2 (q - 1)}\right)}{\sqrt{q - 1} ~~\Gamma \left(\frac{1}{q -1}\right)}
\label{eq:81a}
\end{equation}
\noindent where $\Gamma$ is the Euler $\Gamma$ function and  $1 < q < 3$ in order that the distribution is normalized to one. In this case we have proved that, for $ 1 < q < 5/3$, the function $\rho$ has the following exact expressions:
\begin{equation}
 \rho(q)=\sqrt{\pi} ~\frac{\sqrt{\frac{q - 1}{5 - 3 q}}  ~~\Gamma \left( \frac{3 - q}{2 ( q -1)} \right)}{\Gamma \left(\frac{2 - q}{q - 1} \right)}
\label{eq:8b}
\end{equation}
\noindent We remark that, in the limit $q \rightarrow 1$, the Tsallis distribution becomes the Gauss distribution.
In the specific example of the FPU $\beta$ system, $\theta$ is the mean value of the moduli of differences
\begin{equation}
\xi_{i}=\eta_i-\langle \eta_i \rangle,
\end{equation}
numerically obtained and $\sigma$ the standard deviation:
\begin{equation}
\theta = \frac{\sum |\xi_{i}|}{M}, \qquad \sigma = \sqrt{\frac{\sum \xi_{i}^{2}}{M}}
\label{eq:6}
\end{equation}
\noindent where $M$ is the number of values of $\xi_{i}$. What one expects is that for $\epsilon < \epsilon_{t}$, when the system is stable, $\rho(\epsilon)$ should remain approximately constant. Instead it should change abruptly for $\epsilon > \epsilon_{t}$, when the $\pi$--mode starts to exchange energy with the others modes. For larger and larger values of $\epsilon$, when an equipartition state has been reasonably reached, the parameter $\rho$ should assume again a constant value, characteristic of the distribution of the $\xi_{i}$. For intermediate values of $\epsilon$, a transition between weak and strong chaos should be observed.

\section{Numerical Results}

We describe now our numerical analysis of the FPU $\beta$ model. In order to study numerically the stability of the $\pi$--mode, we utilize the method used in ref \cite{caf}. The equations of the motion in the variables $x_i, p_i$ are integrated by means of a bilateral symplectic algorithm (\cite{Casetti}). We recall that the dynamical properties of the FPU $\beta$ system depend only on the product $\epsilon \beta$, so in all numerical experiments we put $\beta = 1$ and change the value of the energy density without loss of generality. We excite the nonlinear $\pi$--mode at $t=0$, by putting
\begin{equation}
Q(0)=Q_0\neq0, \qquad \dot{Q}(0)=P_0=0.
\end{equation}
From these values, the initial values of $x_i$ and $p_i$ are calculated and the Hamilton equations are integrated in the variables $x_i$ and $p_i$ with an integration step $\Delta t=0.02$. Every $100$ integration steps the quantities \begin{equation}
\eta_i=x_i(t)+x_{i-1}(t), \qquad i=1,\ldots,N
\end{equation}
are calculated. For each value of $\epsilon$, we follow the evolution of the system for a time approximately equal to $10^{6}/\pi$ periods of the corresponding linear normal mode $(T_{N/2}=\pi)$. Longer integration times give qualitatively the same behaviour.

The numerical results show that the dependence of $\rho$ on $\epsilon$ is qualitatively the same for each choice of $\eta_i$ with $32 \leq N \leq 1024$. We shall discuss in detail these results, through an analysis of the case $N  =128$ and $i = 64$. We recall that the value of the energy density for the direct excitation of the $j$--mode ($j < N/2$) by the instability of the mode $N/2$ is given by \cite{poggi}:
\begin{equation}
\epsilon_{ex} = \frac{1}{3} \left( \frac{1}{\sin^{2}{\pi j/N}} -1 \right) \geq  \epsilon_{t}.
\label{eq:9}
\end{equation}
The first mode that becomes unstable is the mode $N/2-1$, when $\epsilon = \epsilon_{t}$.

\begin{figure}[htbp]
\centerline{\includegraphics[width=1.2\textwidth]{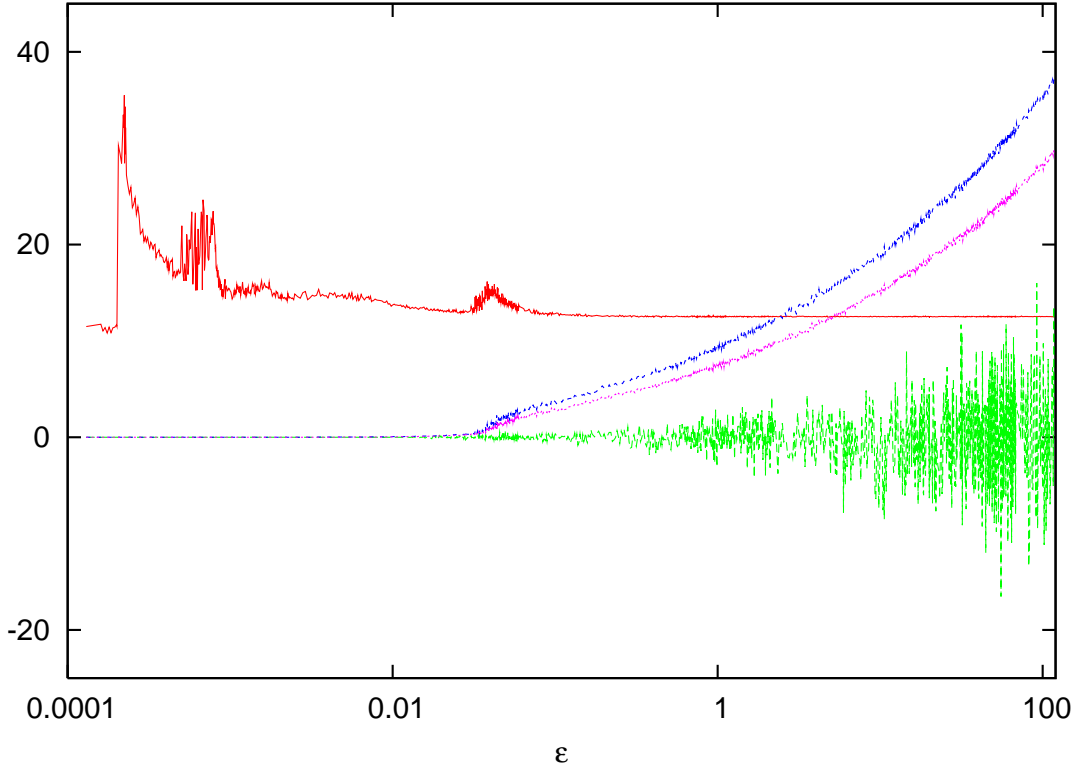}}
\caption{(Color on line). The indicator $\rho$, the first and the second moment $\theta$ and $\sigma$ and $<\eta_{64}>$ vs the energy density $\epsilon$ for $N=128$. For the sake of clarity, the four quantities are rescaled by different numerical factors: $\rho \times 10$ (red), $\sigma \times 2$ (blue), $\theta \times 2 $ (purple) and $<\eta_{64}> \times 500$ (green).}
\label{fig:andrea1}
\end{figure}

In Fig. \ref{fig:andrea1}, the behaviours of $\rho$, $\sigma$, $\theta$ and $<\eta_{64}>$ are shown as a function of $\epsilon$. For the sake of clarity, the four quantities are rescaled by different numerical factors. The global indicator $\rho$ increases abruptly, if $\epsilon$ exceeds the threshold value
\begin{equation}
\epsilon_t = 2.0 \times 10^{-4} \approx \pi^{2}/(3N^{2}).
\end{equation}
Then a rapid decrease of $\rho$, just above $\epsilon_{t}$, is observed with a regular recurrent energy exchange of the $\pi$--mode with the mode $N/2 -1$. This lasts approximately until the mode $N/2 -2$ is directly excited. The excitation energy density of this mode corresponds in the graph of $\rho$ approximately to the presence of a "bush". This energy density can be considered as the beginning of the regime of weak chaos in the system. Moreover, we observe a small peak for $0.01 \leq\epsilon\leq 0.1$. For larger values, $\rho$ is almost independent of $\epsilon$ and reaches the value $\sqrt{\pi/2}$ characteristic of the Gaussian distribution. In this region, the transition to the chaotic behaviour is rapid and the exchange of energy with the other linear modes is complete.

\begin{figure}[htbp]
\centerline{\includegraphics[width=1.2\textwidth]{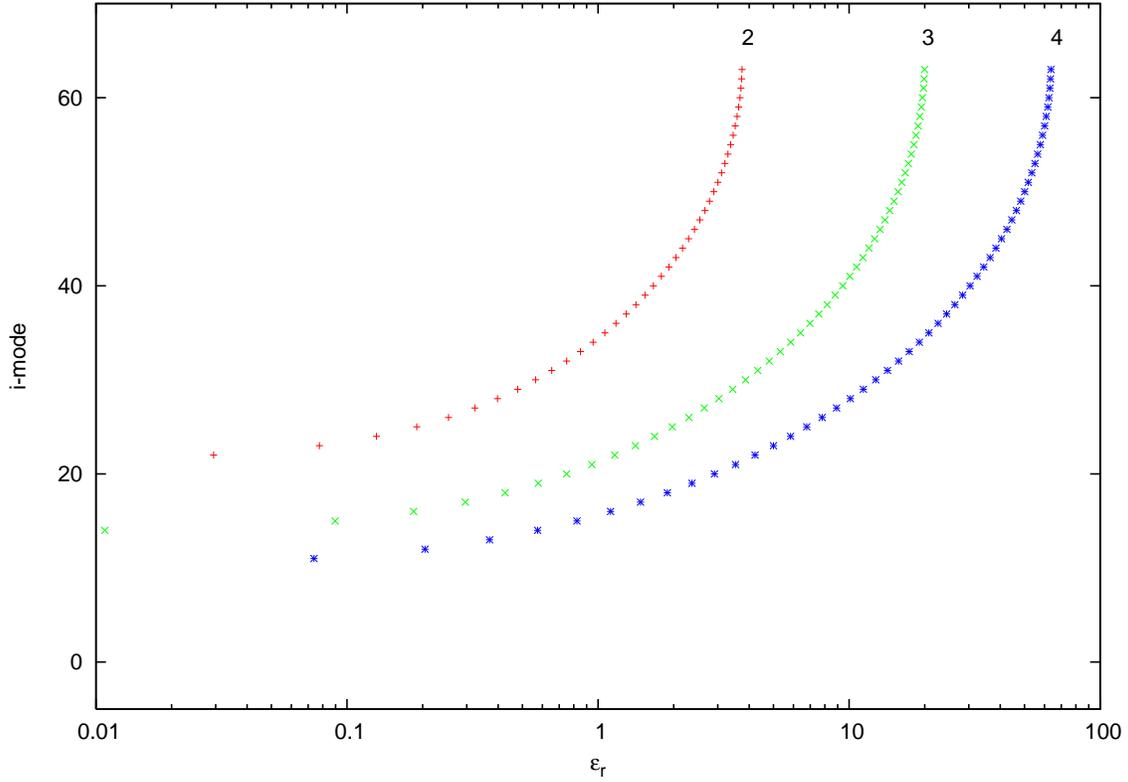}}
\caption{(Color on line). For a direct comparison with the results shown in Fig. \ref{fig:andrea1} and to remark the presence of resonances for energies in correspondence of the small peak in the graph of $\rho$, we report ( see Eq. (\ref{eq:12})) the mode number $i$ as a function of the corresponding resonance energy density $\epsilon_{r}$ for $N =128$ and $m = 2$ (red), $3$ (green) and $4$ (blue).}
\label{fig:andrea2}
\end{figure}
Concerning the behaviour of  $\langle \eta_{64} \rangle$, a significative change is observed  approximately in correspondence of the small peak present in the curve $\rho(\epsilon)$. This change reveals a strong breaking of the symmetry $\eta_{64} = x_{64} + x_{63}=0$ of the $\pi$--mode and marks the transition of the system from weak to strong chaos. The small peak in the plot of $\rho$ could be the consequence of a mechanism of \textit{resonance overlap}. As is well known, the solution of the differential equation for the modal variable $Q$, is given by
\begin{equation}
Q(t) = Q_{0} ~cn (\Omega t;k^{2}),
\end{equation}
where $cn$ is the periodic Jacobi elliptic function with period $T = a K(k)/\Omega$, $K(k)$ is the complete elliptic integral of the first kind and, for $\beta = 1$:
\begin{equation}
k^{2} = \frac{1}{2} \frac{\sqrt{1 + 4 \epsilon} -1}{\sqrt{1 + 4 \epsilon}}, \qquad \Omega^2 = \frac{4}{1 - 2 k^{2}}.
\label{eq:10}
\end{equation}
\noindent One has resonance if the harmonic frequencies $\vec{\omega} = (\omega_{1}, \omega_{2}, \ldots, \omega_{N/2})$, concerning the harmonic term of the Hamiltonian, satisfy the relation
\begin{equation}
\vec{m} \cdot \vec{\omega} = \sum_{i}^{N/2} m_{i} \omega_{i} \approx 0
\end{equation}
where $\vec{m}$ is an array of integers and the $\omega_i$ are given by the formula (\ref{puls}).

Since we excite the $\pi$--mode, we have resonance, in particular,  when $\Omega = m \omega_{i}$ with integer $m > 1$ and for some $\omega_{i}$. From previous relations one obtains for the resonance energy density $\epsilon_{r}$:
\begin{equation}
\epsilon_{r} = \frac{1}{4} \left(m^{4} \sin^{4} {\frac{\pi i}{N}} -1 \right).
\label{eq:12}
\end{equation}
\noindent The resonance is possible for values of $i$ such that $\epsilon_{r} > 0$.

For example, for $m = 2$ one has
\begin{equation}
i = \begin{cases} N/6 \qquad\qquad\qquad if \quad N/6 \quad is \quad integer \\
[N/6] + 1 \qquad\qquad if \quad N/6 \quad is\quad not\quad integer. \label{pol}
\end{cases}
\end{equation}

Here, $[x]$ denotes the smallest integer less than or equal to $x$. Consequently the first linear mode  that goes in resonance with the $\pi$--mode corresponds to $i = 22$, for $\epsilon = 0.0282$. This value marks the rising of the small peak in Fig. \ref{fig:andrea1}. In Figure \ref{fig:andrea2}, the values of {\it i} as a function of the resonance energy densities are reported for $m = 2, 3, 4$.

\begin{figure}[htbp]
\centerline{\includegraphics[width=1.2\textwidth]{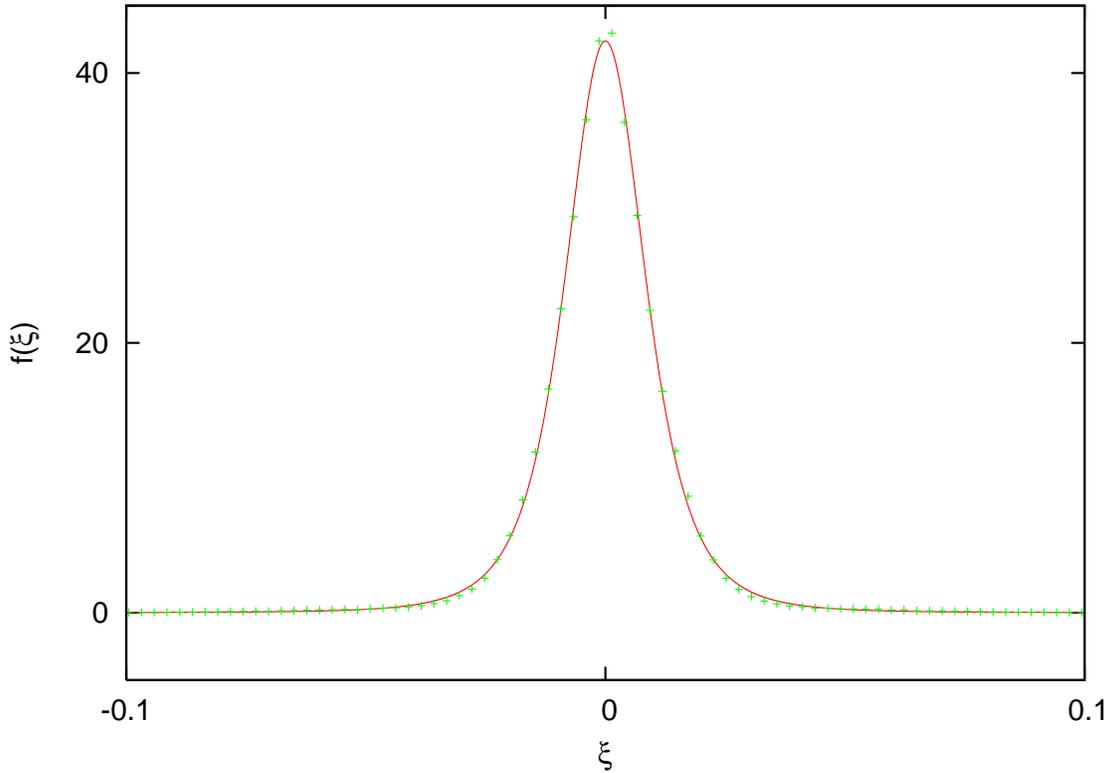}}
\caption{(Color on line). Numerical distribution $f(\xi)$ (green points) of the values of $\xi = \eta_{64} - <\eta_{64}>$ fitted  with  Tsallis distribution (red curve)  for $N = 128$ and $\epsilon = 0.006$.}
\label{fig:andrea3}
\end{figure}

\begin{figure}[htbp]
\centerline{\includegraphics[width=1.2\textwidth]{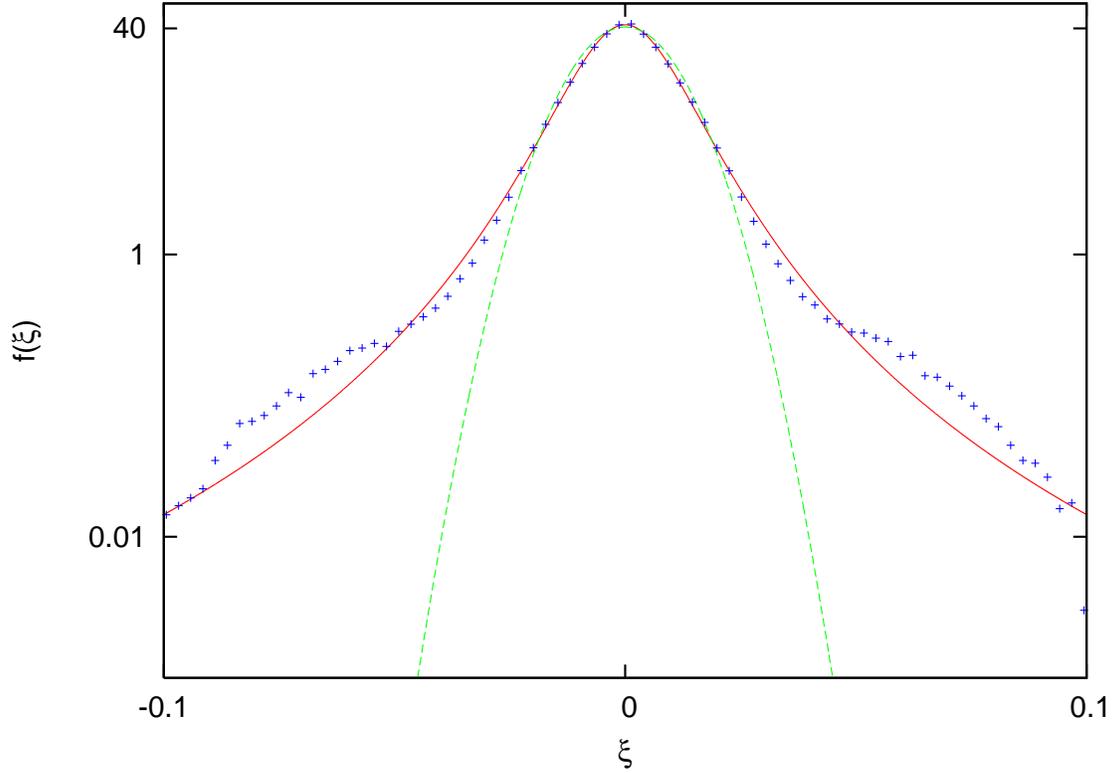}}
\caption{(Color on line). Plot in linear--log scale of the numerical distribution $f(\xi)$ (blue points)  fitted with Tsallis distribution (red) and  Gauss distribution (green) for $N = 128$ and $\epsilon = 0.006$.}
\label{fig:andrea4}
\end{figure}

\begin{figure}[htbp]
\centerline{\includegraphics[width=1.2\textwidth]{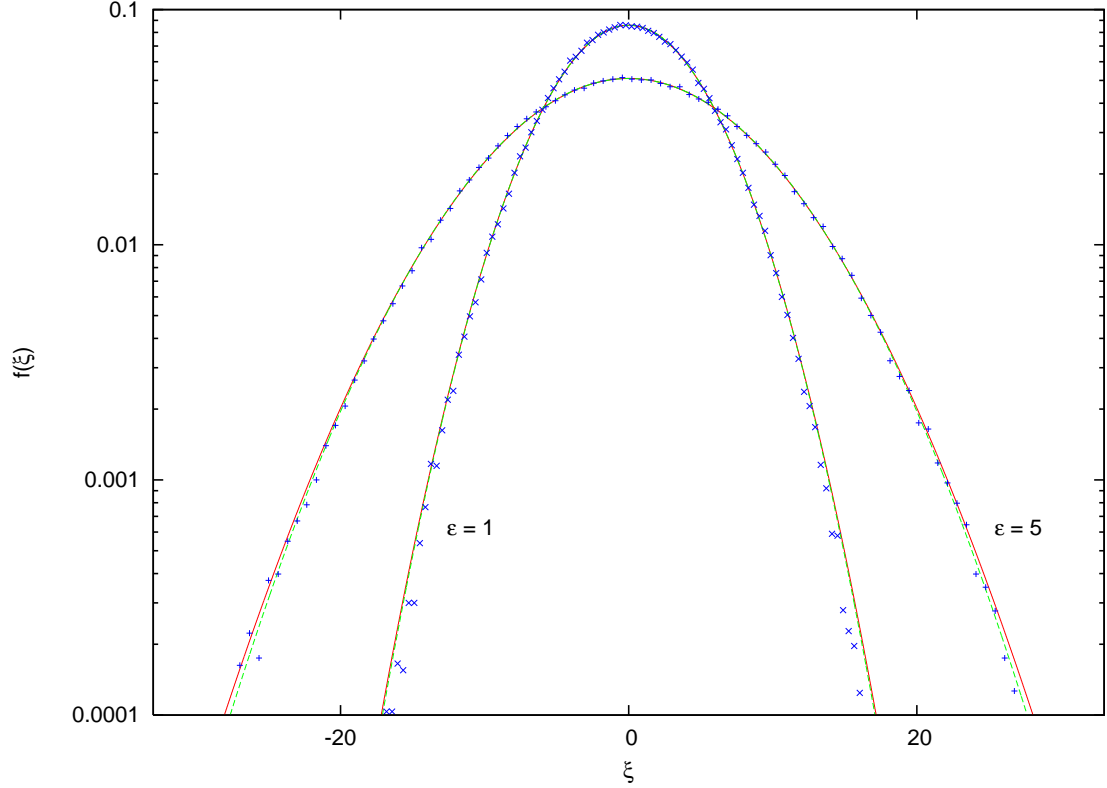}}
\caption{(Color on line). Plot in linear--log scale of the numerical distribution $f(\xi)$ (blue points) fitted with Tsallis distribution (red) and  Gauss distribution (green) for $N = 128$, $\epsilon = 1$ and $\epsilon = 5$. In both cases the Tsallis and Gaussian distributions essentially overlap.}
\label{fig:andrea5}
\end{figure}

Finally, we have analyzed the distribution for values of $\epsilon$ between the "bush" and the small peak in the plot of $\rho$ in Fig. \ref{fig:andrea1}. We have fitted the numerical distribution with a Gaussian and a Tsallis distribution (see Eqs. (\ref{eq:5}), (\ref{eq:81})). It emerges that in the region of weak chaos the numerical distribution is fitted accurately with a Tsallis distribution. A typical fit is shown in Fig. \ref{fig:andrea3} for $\epsilon = 0.006$. The best fit with the Tsallis distribution, with the two parameters $q$ and $a$, gives $q = 1.463$ and $a = 42.380$, with a reduced $\chi^{2} = 0.064$. With this value of $q$ we have, from Eq. (\ref{eq:8b}), a value of $\rho = 1.497$, to be compared with the numerical value $\rho = 1.461$. A fit with the Gaussian distribution with the parameter $a$ gives $a = 40.874$ with a reduced $ \chi^{2} = 0.905$. In Fig. \ref{fig:andrea4} Tsallis, Gaussian and numerical distributions are compared, using a linear--log scale, for  $\epsilon=0.006$. In Fig. \ref{fig:andrea5}, the same distributions are compared for $\epsilon=1$ and $\epsilon=5$. By increasing the energy, the three distributions collapse into the Gaussian one, as we expected. We also get a common value $q=1.01$, that is a signal that we have already reached a region of strong stochasticity.

\section{Open problems}

We have described the evolution of the $\pi$--mode solution of the FPU $\beta$ system, by means of a new indicator of stochasticity. From numerical and analytical results we deduce that, for $\epsilon > \epsilon_{t}$, there are three different regimes in the transition  from a regular to chaotic behaviour. A first KAM--like regime, characterized by a regular and recurrent behaviour, extends approximately from $\epsilon_{t}$ to  the energy at which the mode $N/2 - 2$ is directly excited by the $\pi$--mode. This value of $\epsilon$ corresponds roughly at the appearance of the "bush" in the graph of $\rho$; then a second regime is observed until the small peak (playing the r\^ole of strong sthochasticity threshold) is reached. This is the zone where the  weak chaos dominates. Then, the system enters a regime of strong chaos characterized by the full symmetry breaking of the $\pi$--mode solution. From our analysis it emerges with a good evidence that the regime of weak chaos is described by the Tsallis distribution.

It would be interesting to extend this analysis to other exact solutions. It is a completely open question to ascertain whether, after a sufficiently long time, the weakly chaotic regime here described would collapse into a fully chaotic one. This aspect can be connected with the recent investigation on the \textit{metastability scenario} for the FPU problem \cite{Fucito}, \cite{BCGG}, \cite{BP}.

Finally, it would be important to analyze, from the perspective of nonextensive thermostatistics, also the case of the Fermi--Pasta--Ulam system with fixed boundary conditions (for a recent study, see \cite{Bountis}).

\section*{Acknowledgments}

The authors are grateful to Prof. Tsallis for his attention to this work and a careful reading of the manuscript. P. T. thanks him gratefully for many useful discussions. The research of M. Leo and R. A. Leo has been supported by MIUR, Italy.  The research of P. T. has been supported by the grant FIS2008--00260, Ministerio de Ciencia e Innovaci\'on, Spain. P. T. also wishes to thank the Dipartimento di Fisica, Universit\`a del Salento, for warm hospitality and financial support.

\end{document}